\documentclass{jpsj2}
\begin{document}

\title{Aharonov-Bohm oscillation of magnetization in two-dimensional corbino disk
}

\author{Katsumi \textsc{Itahashi,}$^{1}$
Keita \textsc{Kishigi,}$^{1,}$\thanks{E-mail address: kishigi@educ.kumamoto-u.ac.jp} 
 and Yasumasa \textsc{Hasegawa}$^{2,}$\thanks{E-mail address: hasegawa@sci.u-hyogo.ac.jp}}

\inst{$^{1}$ Faculty of Education, Kumamoto University, Kurokami 2-40-1, 
Kumamoto 860-8555, Japan \\
$^{2}$ Department of Material Science, Graduate School of Material Science, 
University of Hyogo, Hyogo 678-1297, Japan \\
%
}


%

\date{\today}

\abst{
We numerically calculate the magnetization by applying the magnetic field ($B$) perpendicular to two-dimensional corbino disk system without electron spin. We obtain that the period of the Aharonov-Bohm (AB) oscillations of magnetization $[M(B)]$ exhibit $\phi_0$, $\phi_0/2$ and almost $\phi_0/3$ as a function of $\phi$, depending on numbers of electrons ($N$), where $\phi_0=\frac{hc}{e}$ is a unit flux and 
$\phi$ is the magnetic flux through the hollow of the disk. 
The oscillations of $M(B)$ are classified into five patterns at $1\leq N\leq 27$ investigated in this study. 
These are expected to be observed in the two-dimensional 
semiconductor corbino disk with the hollow radius of about 10 nm and small electron numbers at the high magnetic field.
}

\kword{Aharonov-Bohm oscillation, two-dimensional corbino disk, magnetization}
%
\maketitle

\section{Introduction}

The Aharonov-Bohm (AB) oscillation\cite{AB,Byers} is the phenomena that the current in the rings oscillates as a function of magnetic field ($B$) with the period, $\phi_0$, where $\phi_0=\frac{\hbar c}{2\pi e}\simeq4.14\times10^{-15}$ T$\cdot$m$^2$ is a unit flux and $\phi$ is the magnetic flux through the rings. This is due to the phase difference of the wave function given by the vector potential. The AB oscillation with the saw-tooth shape and the period of $\phi_0$ is also predicted in 
the persistent current for the spinless case\cite{Buttiker,Cheung,Bles} in the one-dimensional metal rings. If the spin is considered\cite{Loss}, although the period is $\phi_0$ for the numbers of electron $(N$) is even, the period is $\phi_0/2$ for odd $N$, which is originated in the energy band splitting by the spin. Experimentally, magnetization caused by the persistent current with the periods of $\phi_0$ and $\phi_0/2$ as a function of $B$ has been observed in copper\cite{Levy}, normal metal\cite{Chand} and semiconductor\cite{Mailly} rings. 

\begin{figure}[bt]
\begin{center}
\includegraphics[width=0.3\textwidth]{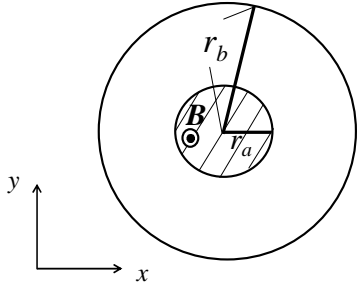}
\end{center}
\caption{
Schematic figure of the two-dimensional corbino disk system.
}\label{fig1}
\end{figure}

The self-organized ringlike semiconductor nanostructures are fabricated by InAs/GaAs\cite{Lorke} and GaSb/GaAs\cite{Timm}. 
In the ring of InAs/GaAs\cite{Lorke}, the height is about 2 nm, 
the inner radius $(r_a$) is about 10 nm and the outer radius $(r_b)$ are about from 30 nm to 70 nm. The value of $r_b/r_a$
is much larger than 1. Therefore, we consider these semiconductor rings as the two-dimensional corbino disk system seen in Fig. \ref{fig1} not the ideal one-dimensional rings. The AB oscillations in the persistent currents carried by single- and 
few-particle states are 
observed in InAs/GaAs quantum rings\cite{Kleemans2007}. 

Using the two-dimensional tight-binding model for graphene\cite{Igor}, for the hexagonal and trigonal monolayer rings and the hexagonal bilayer rings with zigzag boundary conditions it has been shown that the saw-tooth AB oscillation of magnetization [$M(B)$] has the periods of $\phi_0$ and $\phi_0/2$ for even and odd $N$, respectively. 
In the oscillation with the period of $\phi_0$, the wave forms of some types such as the normal, pinched, rounded and asymmetric rounded shapes are seen. In the hexagonal bilayer rings 
case, since some energy levels near the Fermi energy are located in the narrow energy region, the wave form of the oscillation with the period of $\phi_0$ becomes complicated.

Recently, a system with cylindrical hollow structures has been possible from semiconductor nanowire structures by using gold catalyst particles\cite{JWW,Qian}.
Gladilin {\it et al.} \cite{Gladilin} theoretically study the energy spectrum and the magnetization of quasi-two-dimensional (quasi-2D) cylindrical nanoshells. 
They have shown that for the non-interacting case 
the phase and period of $M(B)$ are independent of $N$ within rather wide $N$ ranges. They focus on the quasi-two-dimensionality and 
have never been calculated for the two-dimensional corbino disk system. 


In this paper, 
when the spin is ignored in the two-dimensional corbino disk system, it is shown that the periods of the saw-tooth oscillation of the magnetization become $\phi_0$ or $\phi_0/2$ or almost $\phi_0/3$ when $N$ is changed. The existence of the period of almost $\phi_0/3$ in the magnetization has never been predicted theoretically before, which is caused by the crossing of the energy levels belonging to $n=0$, $n=1$ and $n=2$ near the Fermi energy, where $n$ is the principal quantum number ($n=0, 1, 2 ,\cdots$).

\begin{figure}[bt]
\begin{center}
\includegraphics[width=0.58\textwidth]{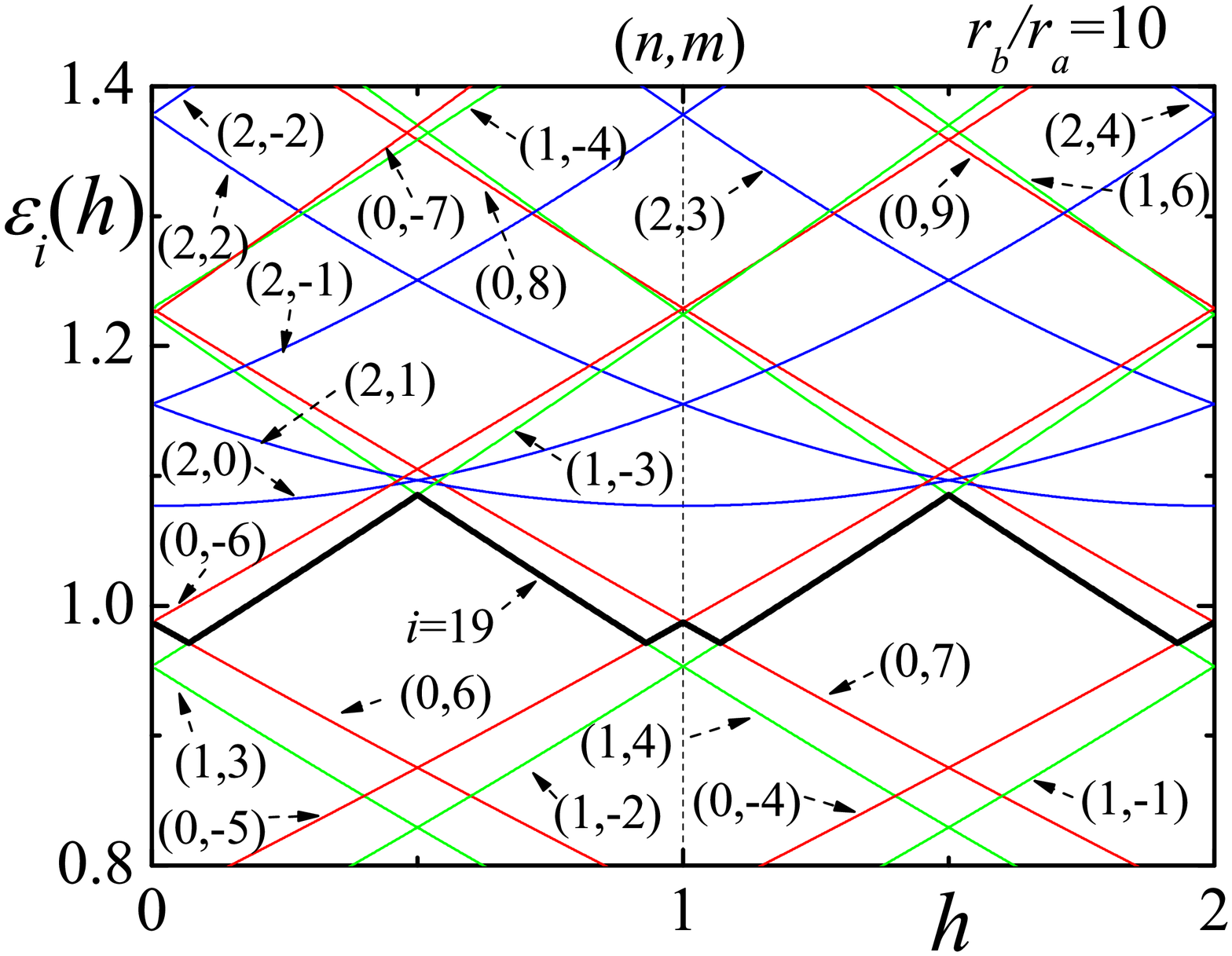}
\includegraphics[width=0.58\textwidth]{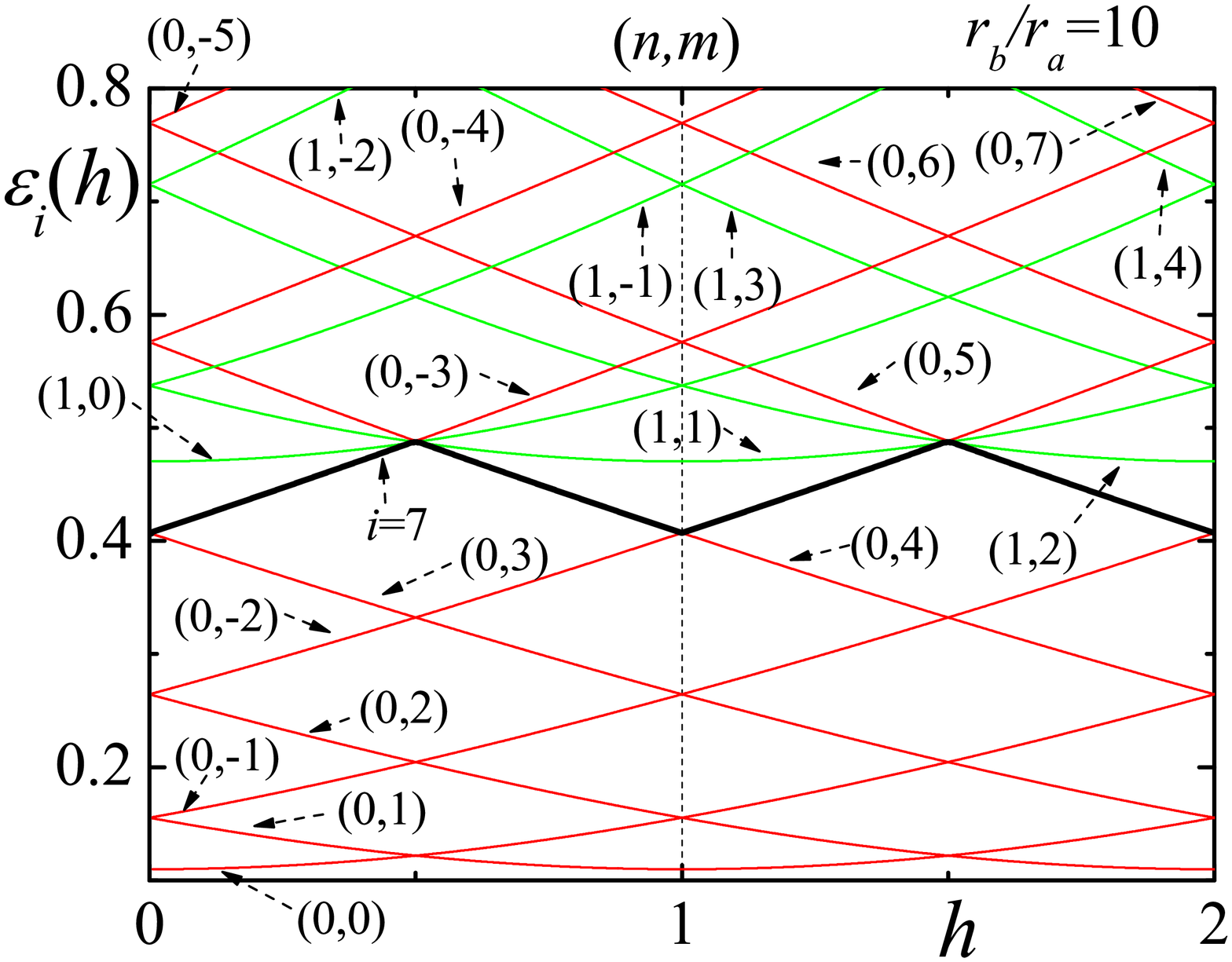}
\end{center}
\caption{
$\varepsilon_{i}(h)$ as a function of $h$ for $r_b/r_a=10$, where $(n, m)$ indicate the quantum numbers. Red lines, green lines and blue lines are for the energy levels belonging to $n=0$, 1 and 2, respectively. 
A black thick solid line in an upper figure and that in a lower figure are for $\varepsilon_{i}(h)$ at $i=19$ and $i=7$, respectively.
}\label{fig2}
\end{figure}

\begin{figure}[bt]
\includegraphics[width=0.5\textwidth]{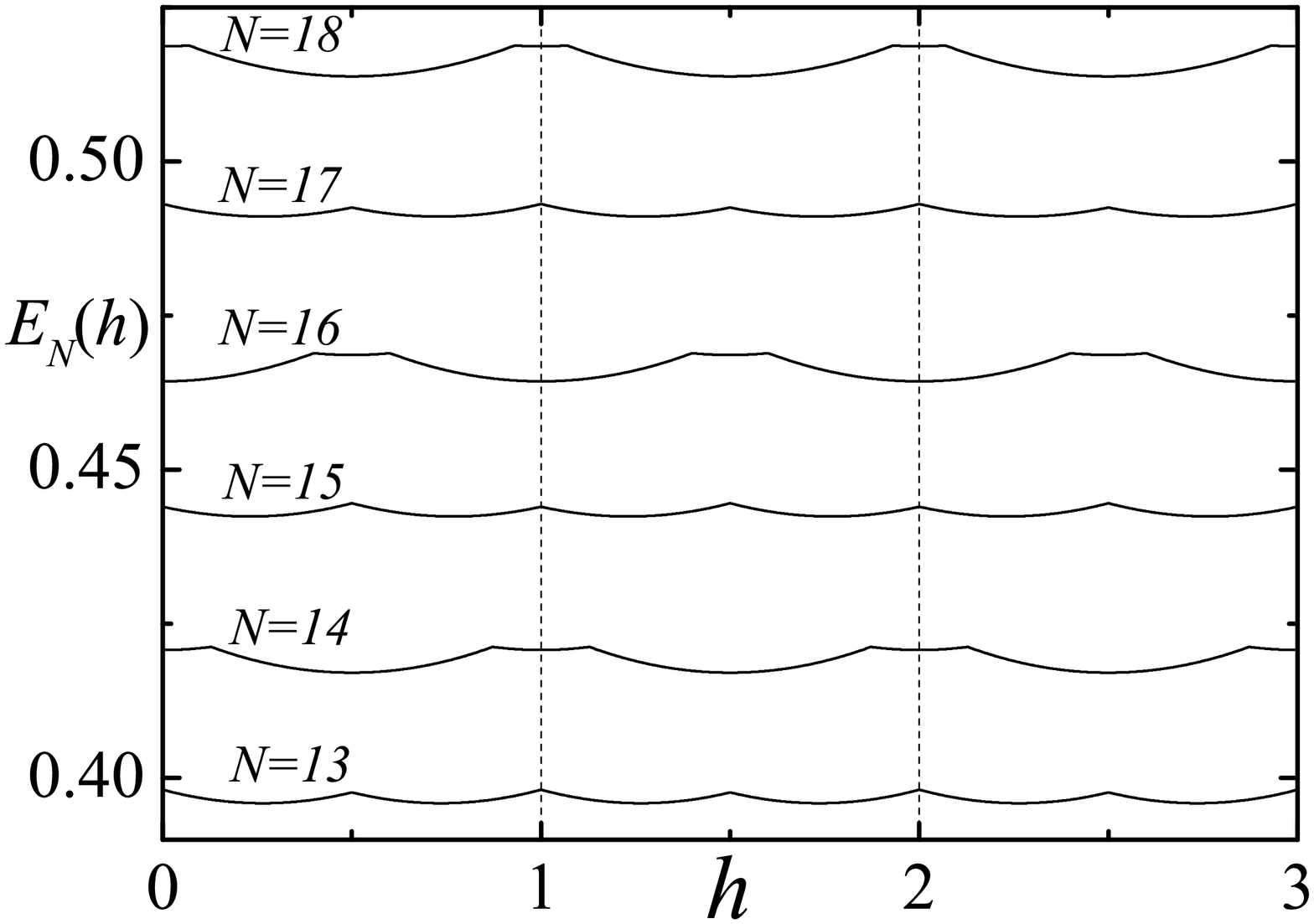}
\includegraphics[width=0.5\textwidth]{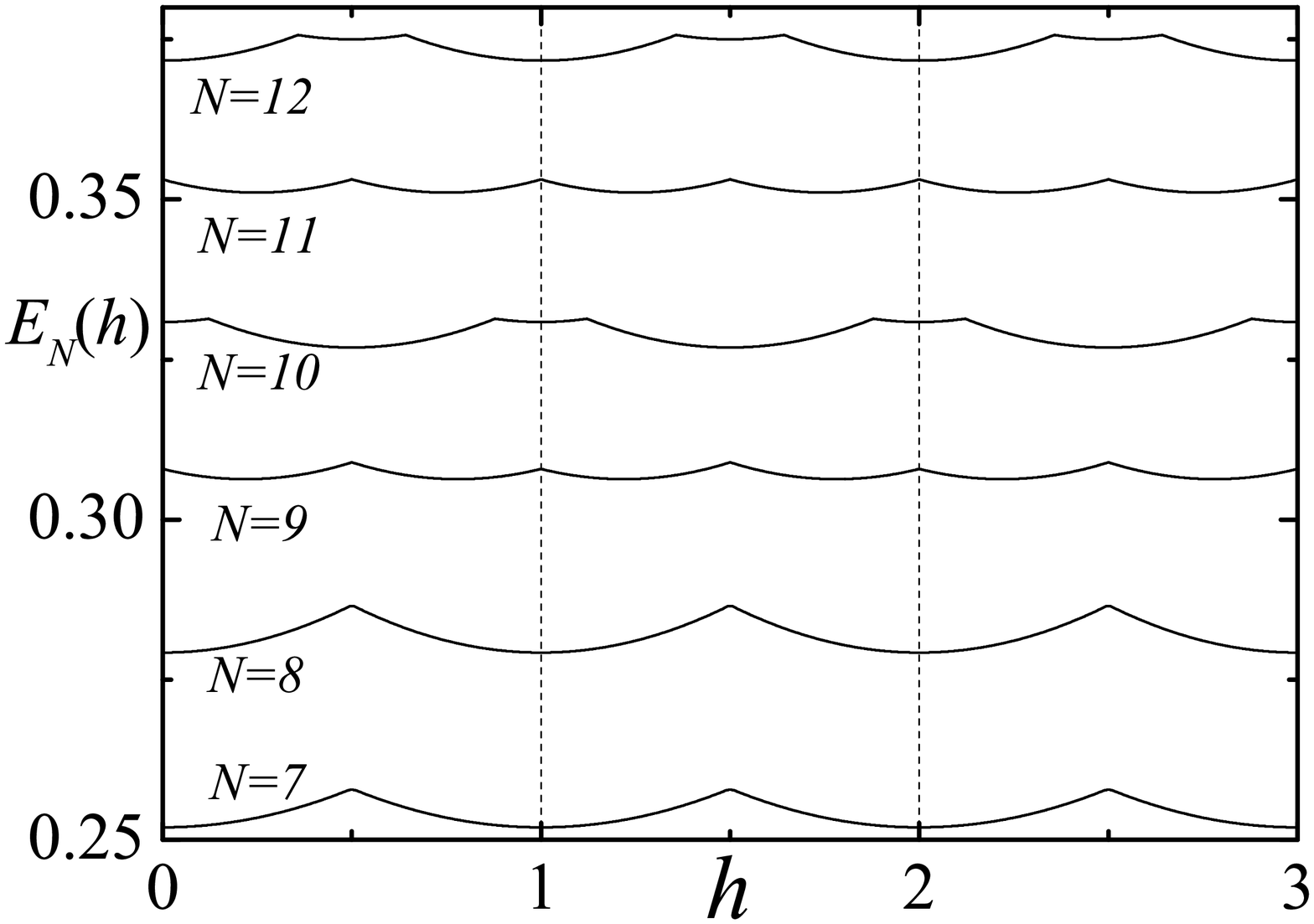}
\includegraphics[width=0.5\textwidth]{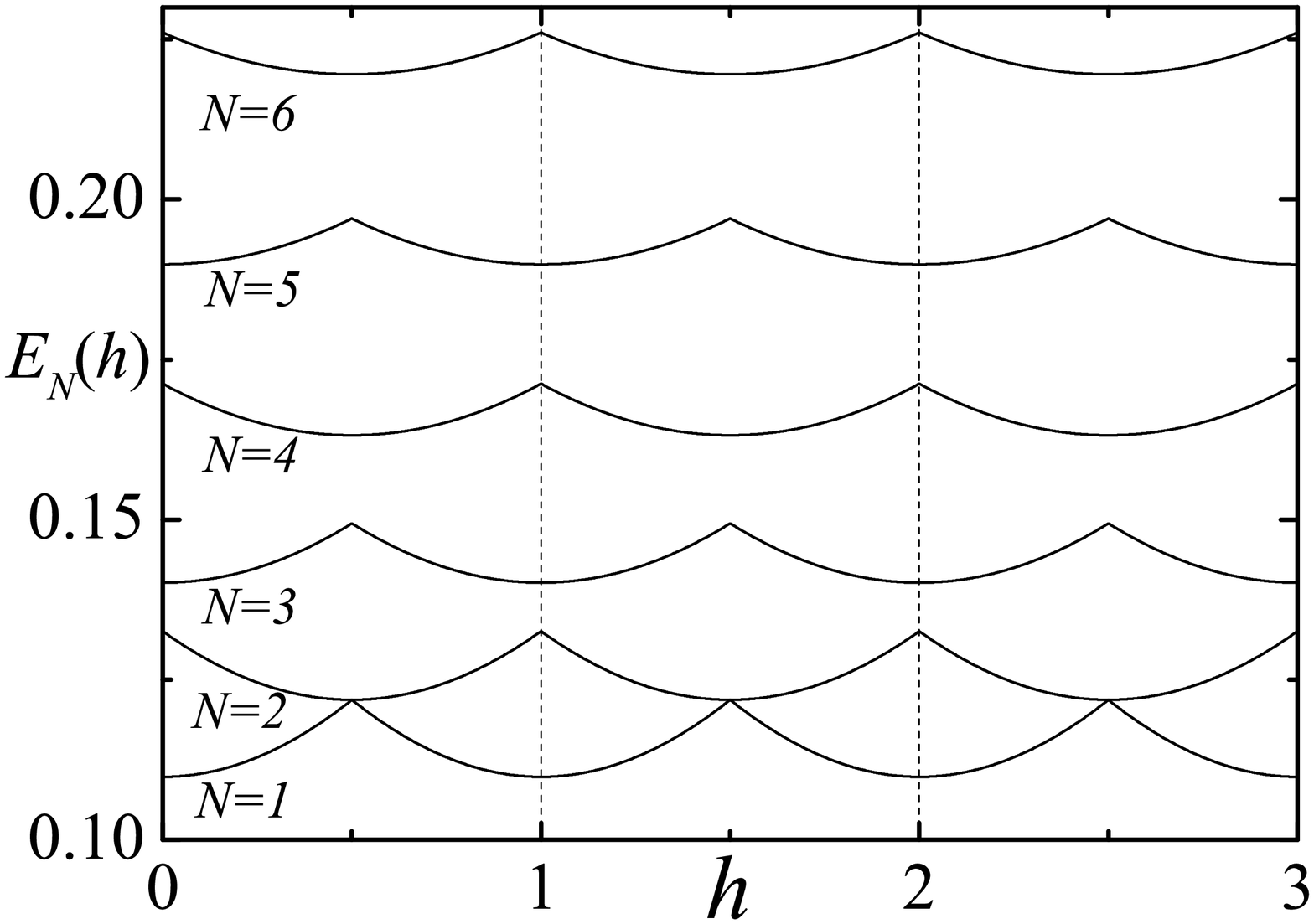}
\caption{
$E_N(h)$ as a function of $h$ for $r_b/r_a=10$ at $1\leq N\leq 18$. 
}\label{fig3}
\end{figure}

\begin{figure}[bt]
\includegraphics[width=0.5\textwidth]{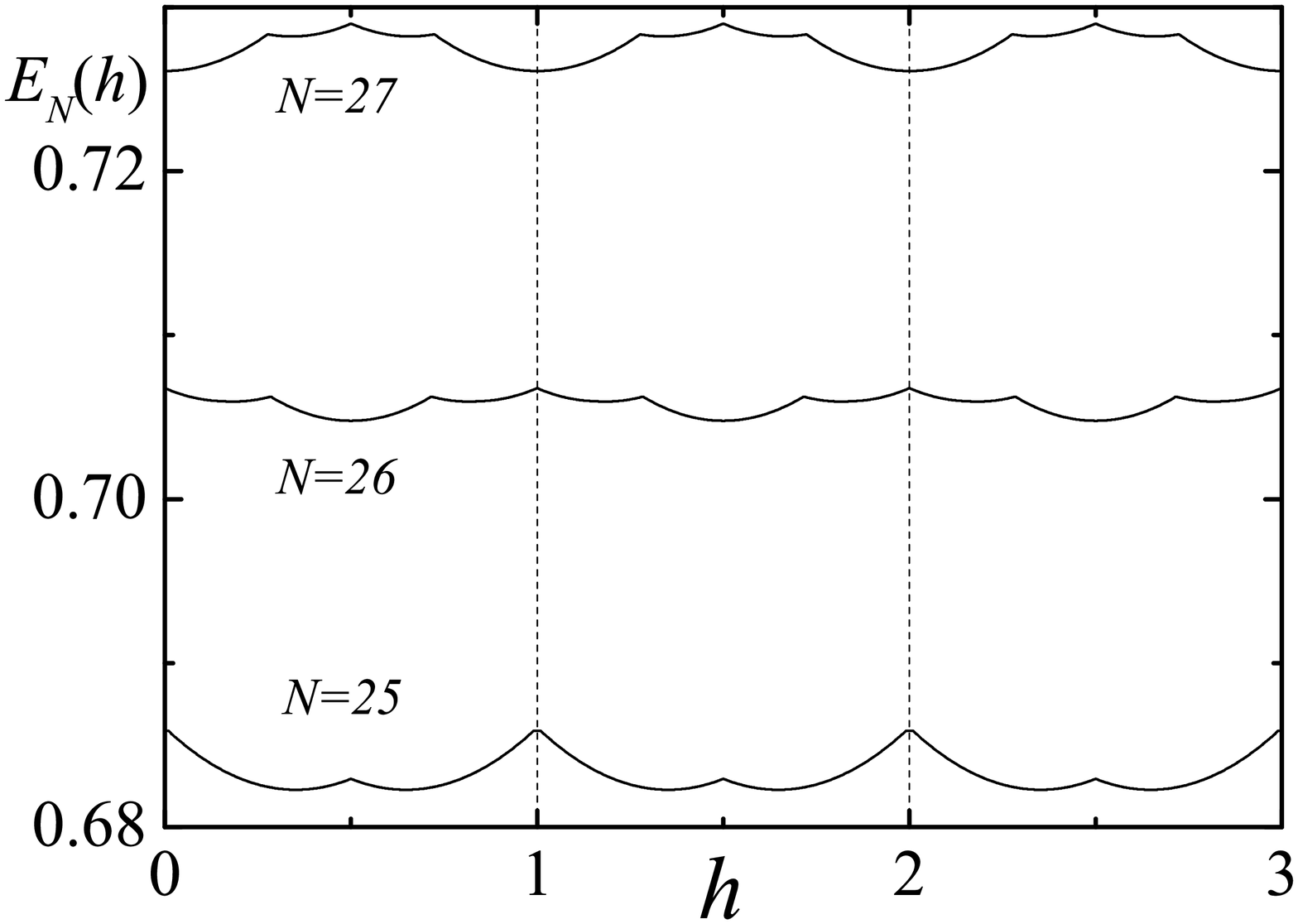}
\includegraphics[width=0.5\textwidth]{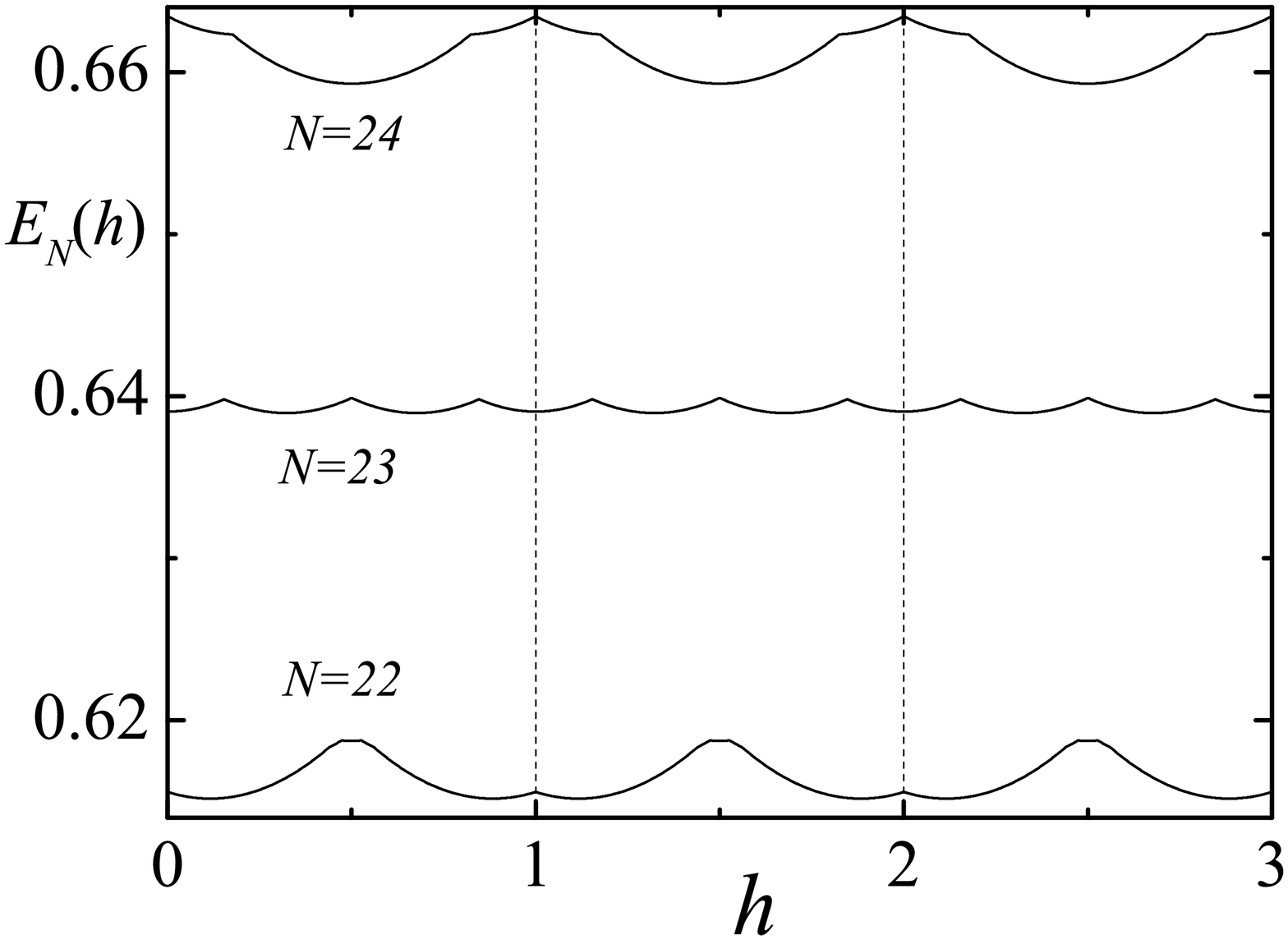}
\includegraphics[width=0.5\textwidth]{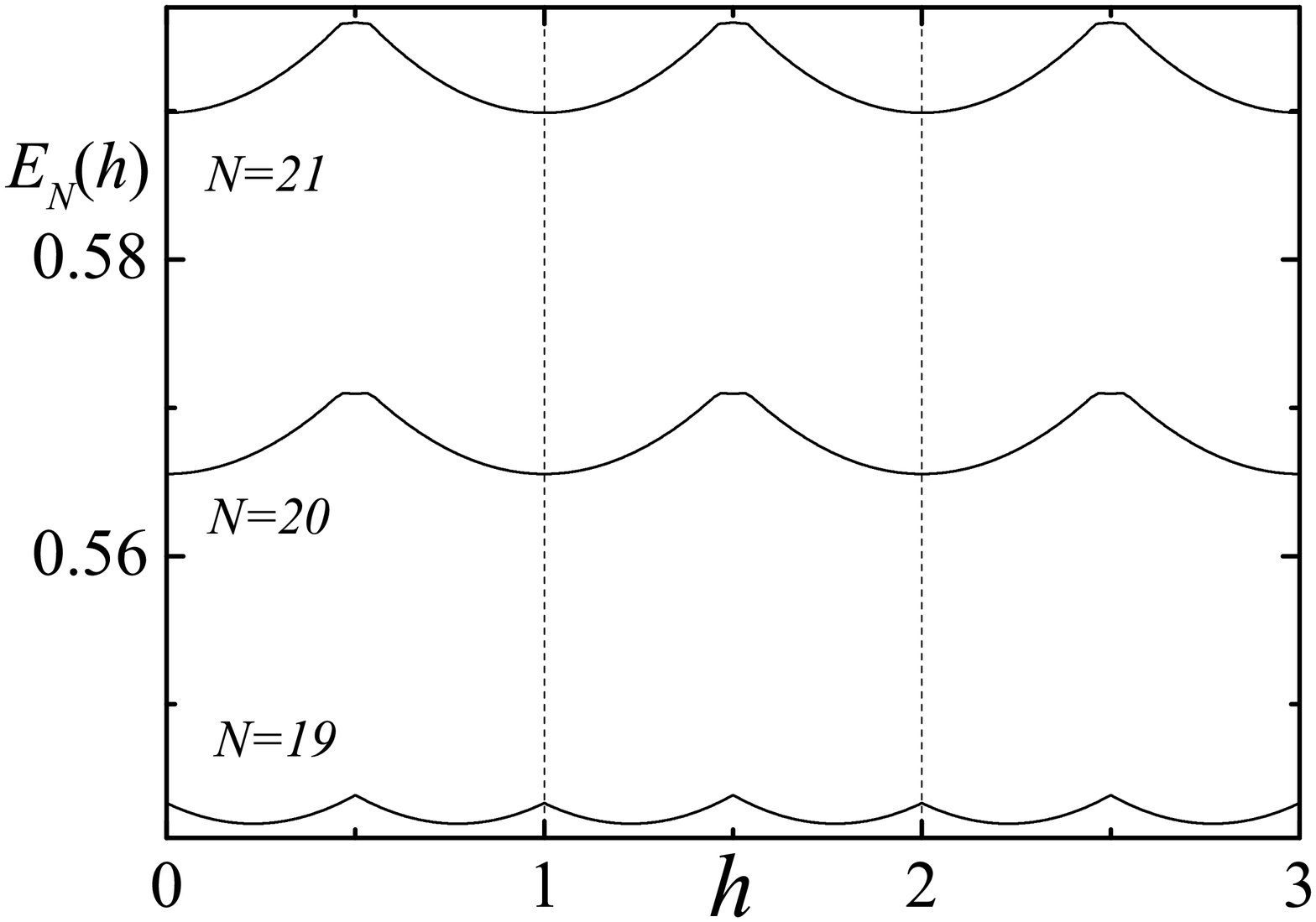}
\caption{
$E_N(h)$ as a function of $h$ for $r_b/r_a=10$ at $19\leq N\leq 27$. 
}\label{fig4}
\end{figure}

\begin{figure}[bt]
\includegraphics[width=0.5\textwidth]{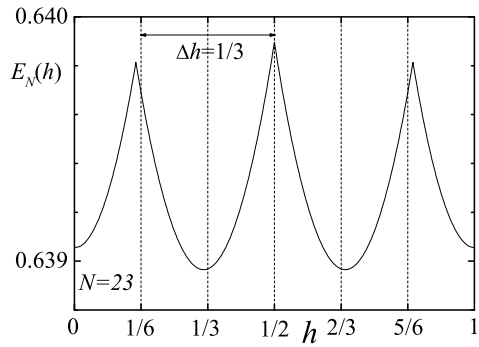}
\caption{
$E_N(h)$ as a function of $h$ for $r_b/r_a=10$ at $N=23$. 
}\label{fig5}
\end{figure}


\section{Model}

\begin{figure}[bt]
\begin{center}
\includegraphics[width=0.37\textwidth]{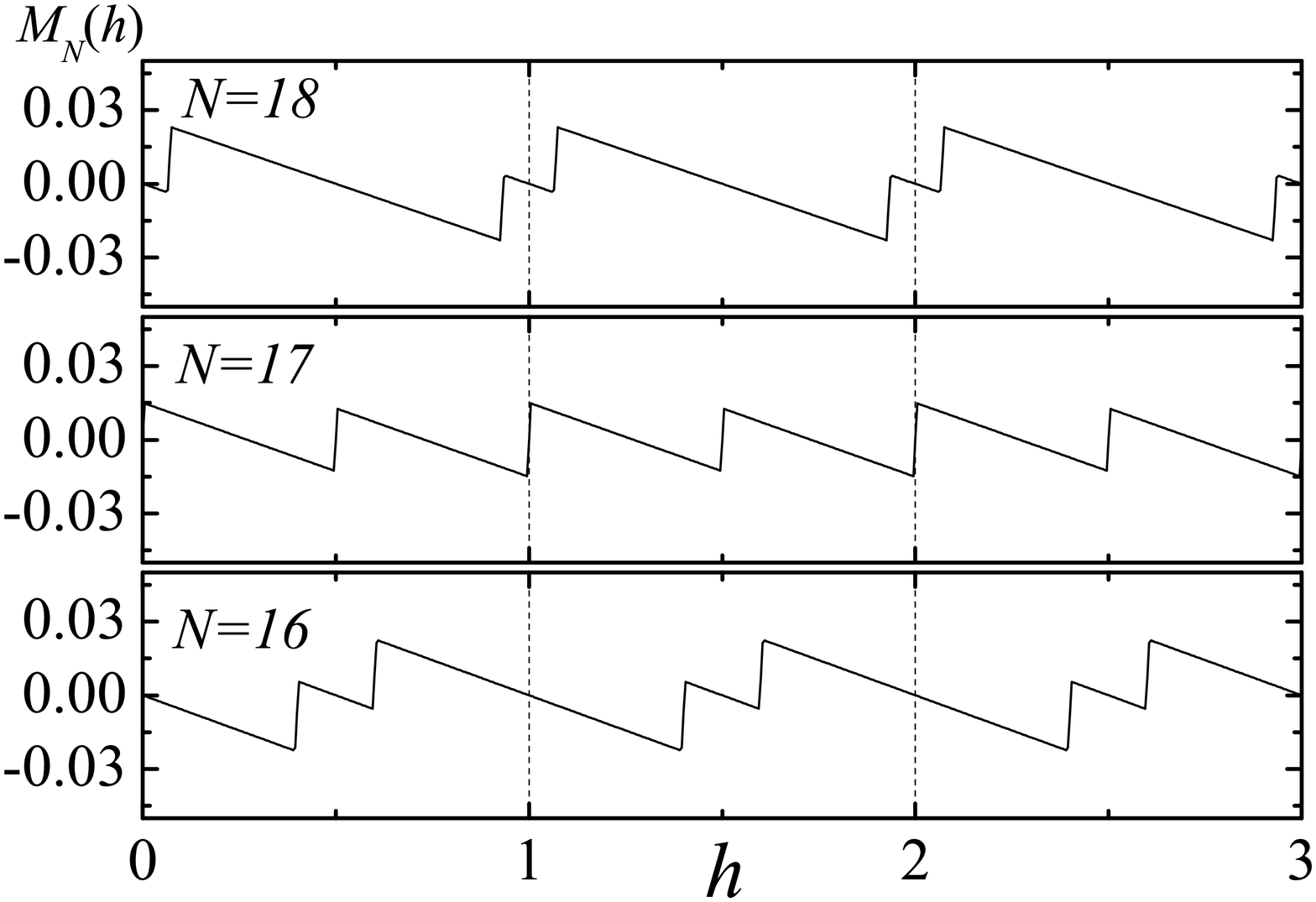}
\includegraphics[width=0.37\textwidth]{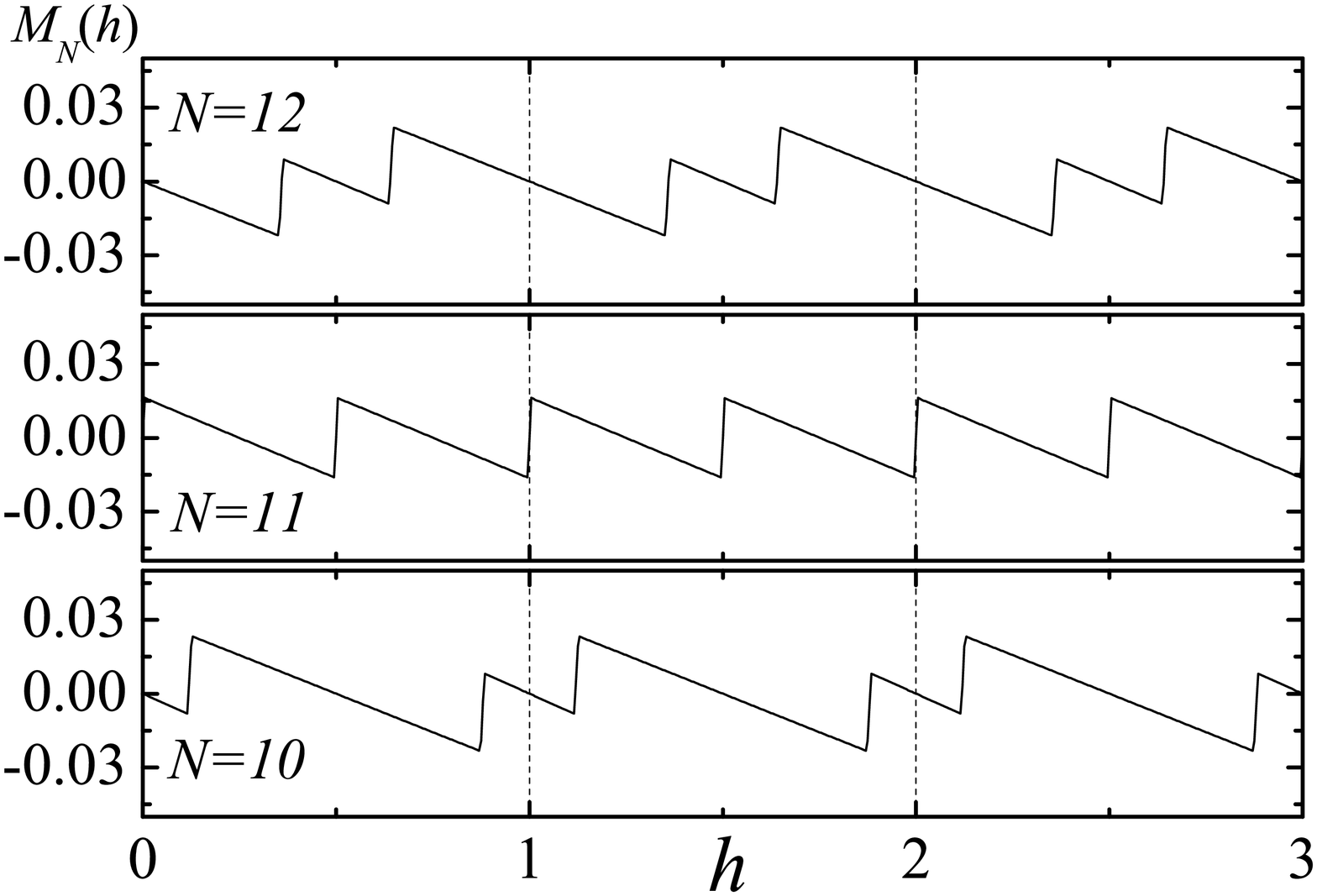}
\includegraphics[width=0.37\textwidth]{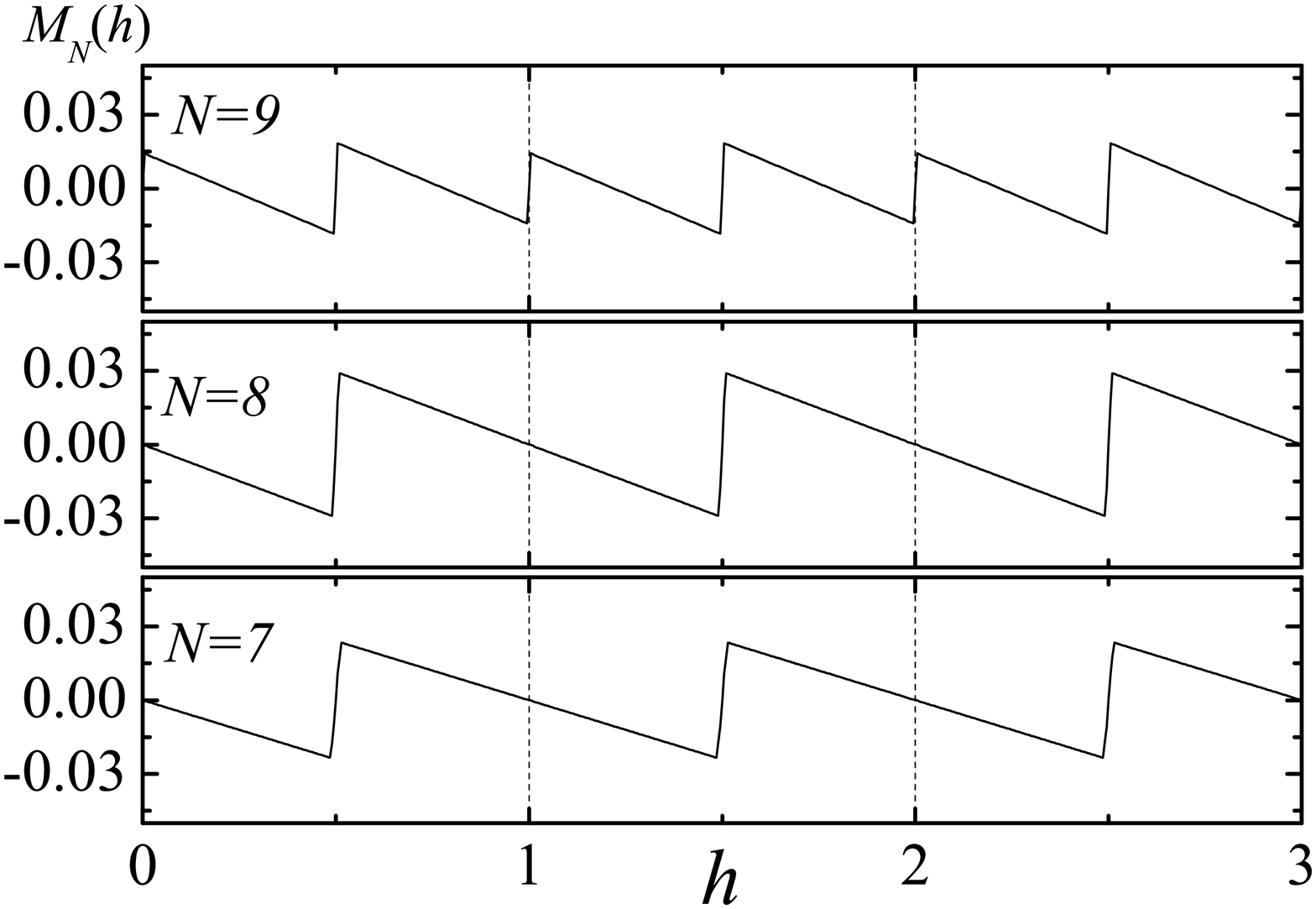}
\includegraphics[width=0.37\textwidth]{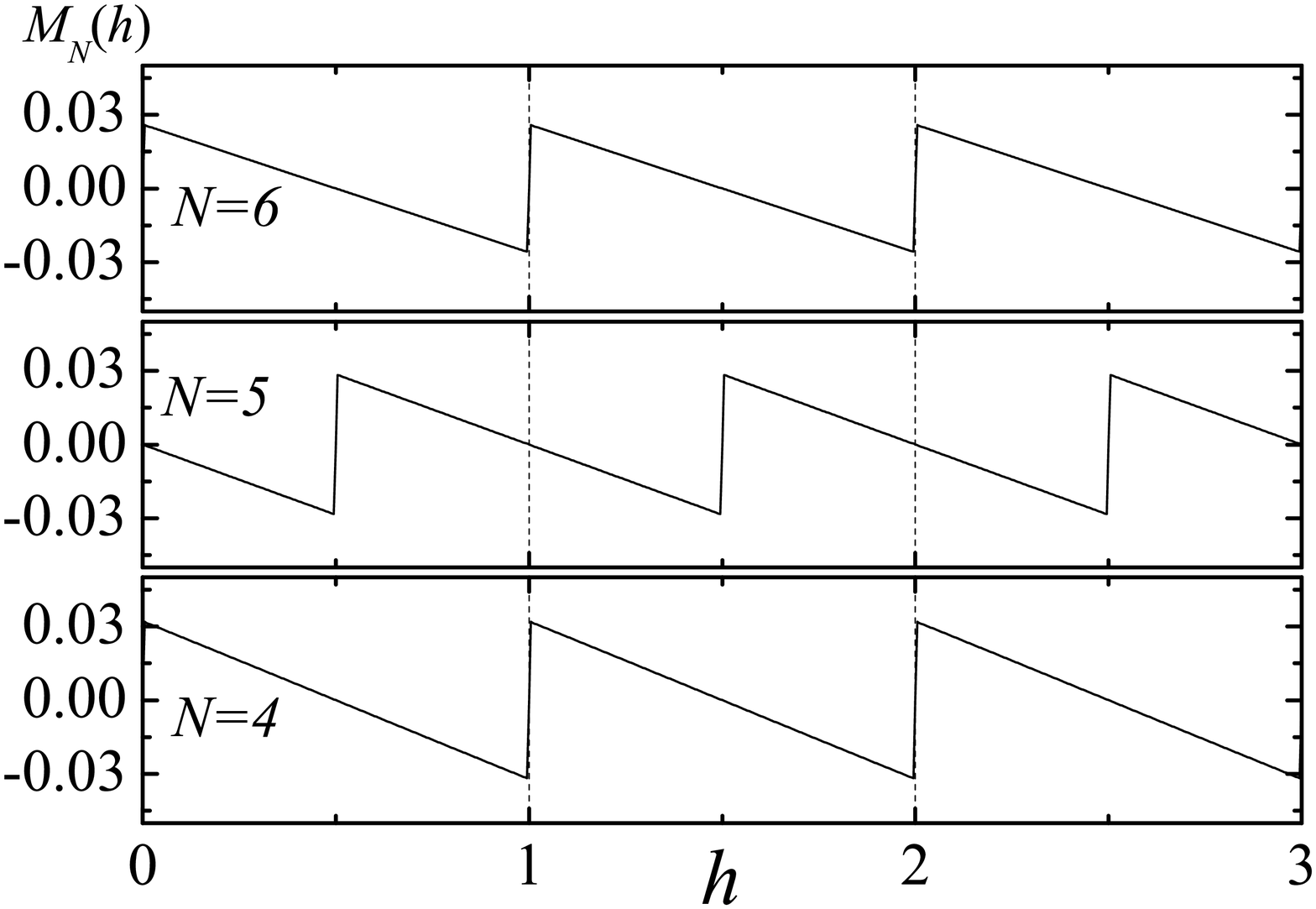}
\includegraphics[width=0.37\textwidth]{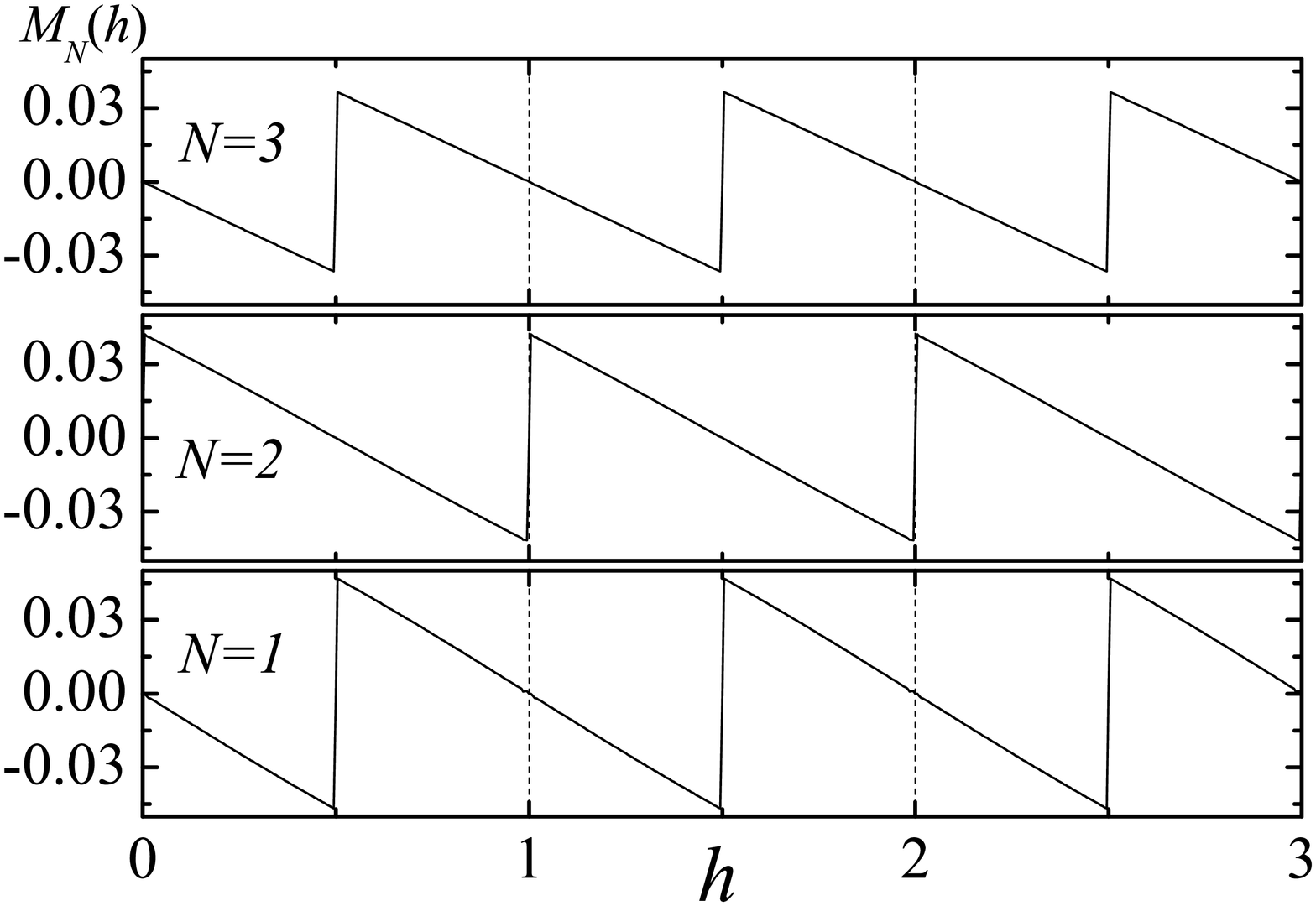}
\end{center}
\caption{
$M_N(h)$ as a function of $h$. Pattern (a): $N=$2, 4 and 6. 
Pattern (b): $N=1, 3$, 5 and 7. Pattern (c): $N=$9, 11, 13 and 17. Pattern (d): $N=$8, 10, 12, 16 and 18.
}\label{fig6}
\end{figure}

\begin{figure}[bt]
\begin{center}
\includegraphics[width=0.37\textwidth]{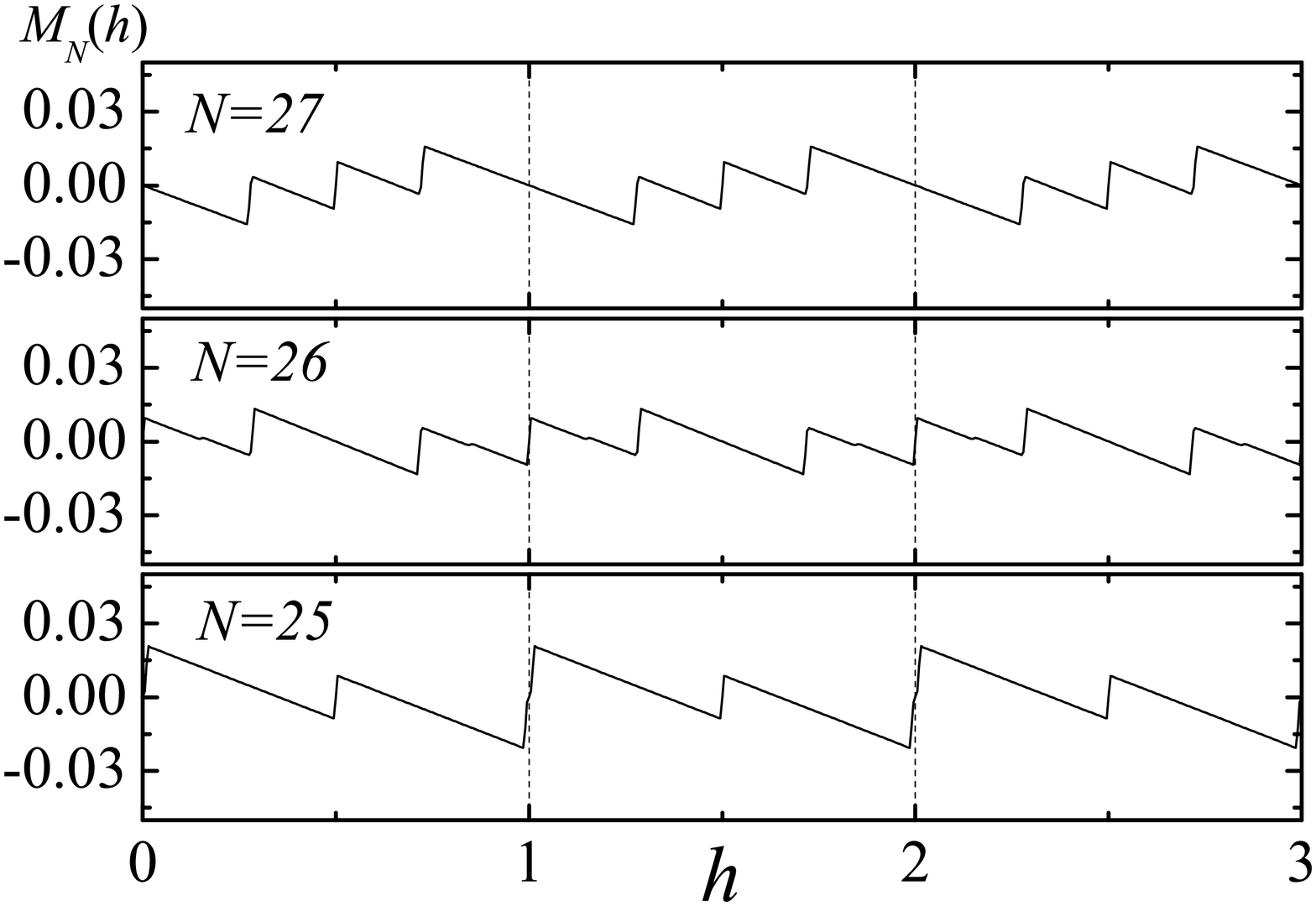}
\includegraphics[width=0.37\textwidth]{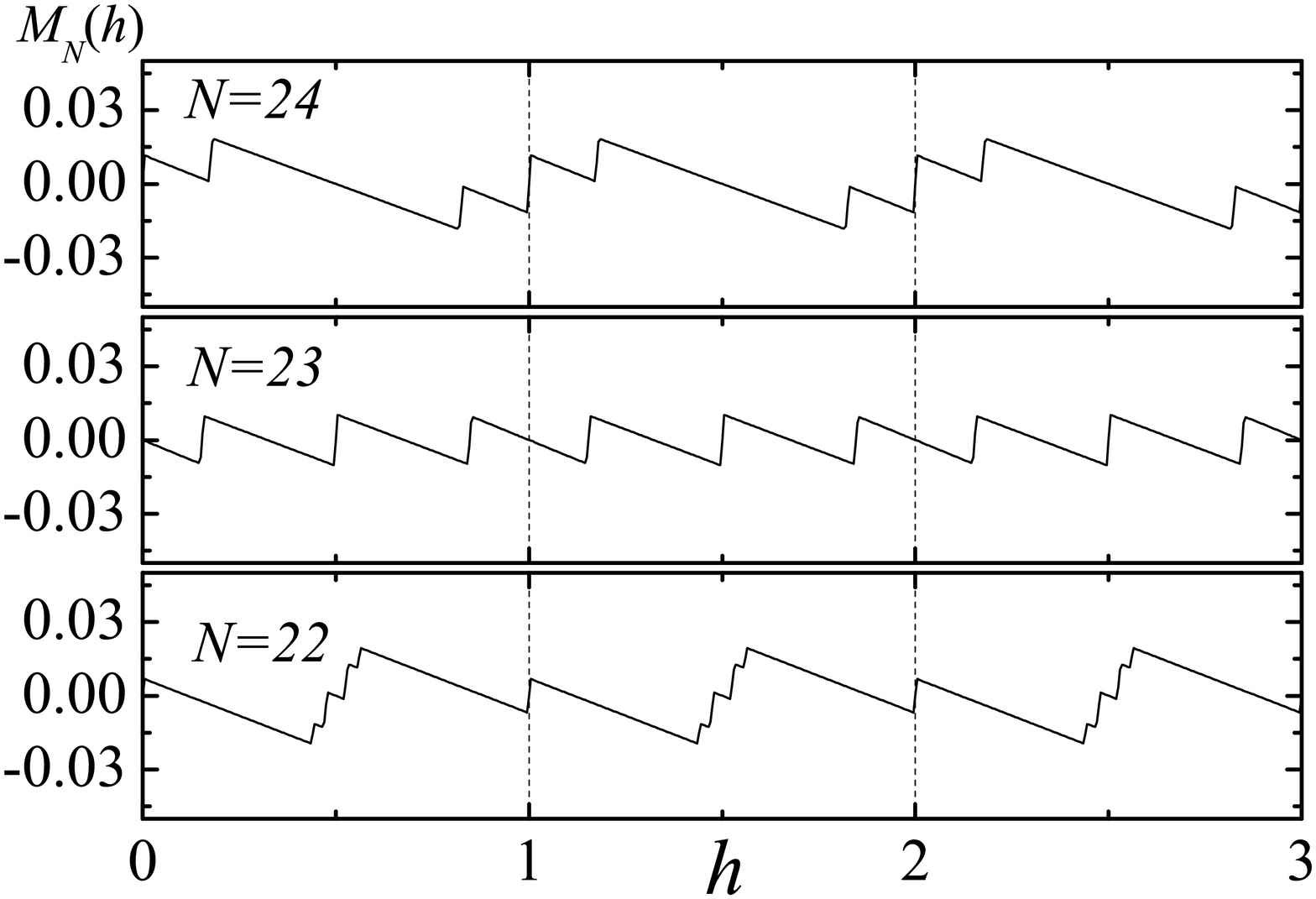}
\includegraphics[width=0.37\textwidth]{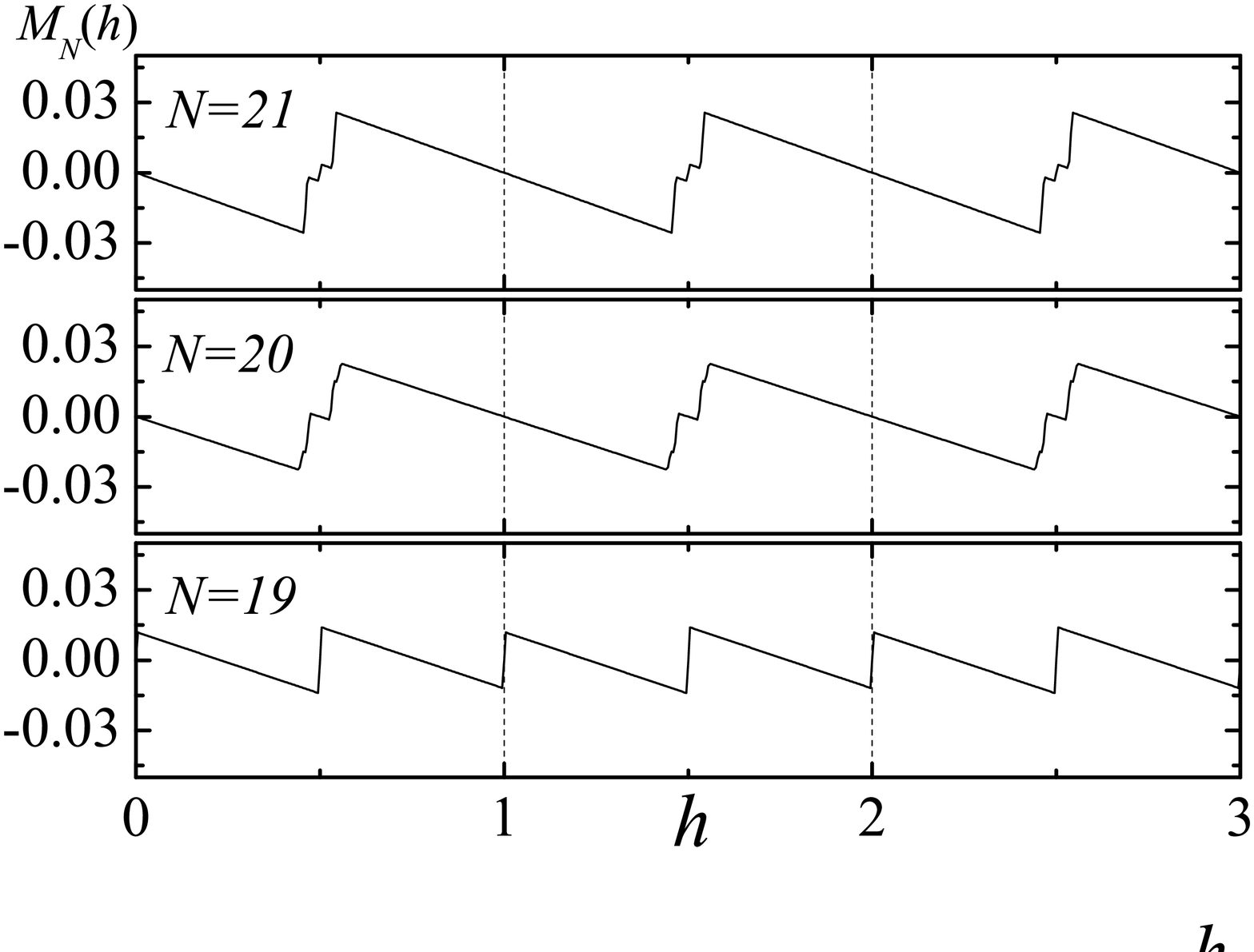}
\end{center}
\caption{
$M_N(h)$ as a function of $h$. Pattern (c): $N=$19 and 25. 
Pattern (d): $N=$20, 21, 22, 24, 26 and 27. Pattern (e): $N=$23. 
}\label{fig7}
\end{figure}

We consider the spinless two-dimensional corbino disk system as seen in Fig. \ref{fig1}, where electrons are confined at the region of $r_a\leq r\leq r_b$. For simplicity we neglect the electron interactions. 
When $r_b/r_a=1$, the system becomes the one-dimensional ring. The magnetic field ($B$) is applied perpendicular to the disk. 
We use polar coordinates ($r, \theta$) and take the gauge 
for the vector potential as follows;
\begin{eqnarray}
\mathbf{A}=(0,\frac{\phi}{2\pi r}),
\label{eq:2.1.6}
\end{eqnarray}
where $\phi=\pi r_a^2B$ is the magnetic flux through the hollow. 
The Schrodinger equation for an electron wave function $(\Psi)$ is given by 
\begin{eqnarray}
&&-\frac{\hbar^{2}}{2m_e}
\bigg[\frac{1}{r}\frac{\partial}
{\partial r}\bigg(r\frac{\partial}{\partial r}\bigg)
+\frac{1}{r^2}\bigg(
\frac{\partial}{\partial \theta}
-ih\bigg)^2\bigg]\Psi (r,\theta) \nonumber \\
&&=\varepsilon_{}(h)\Psi (r,\theta),
\label{eq:2.1.8}
\end{eqnarray}
where $m_e$ is the mass of an electron and $h=\phi/\phi_0$. 
Hereafter, we use $h$ as the unit of the strength of the magnetic field. 
When we choose $r_a=10$nm, $h\simeq 1$ means $B\simeq 1.32\times10^{}$ T. The form of wave function is written by 
\begin{eqnarray}
\Psi (r,\theta)=e^{im\theta}R(r), 
\label{wave} 
\end{eqnarray}
where $m=0, \pm 1, \pm 2, \cdots$ are quantum numbers for $\theta$. 
By putting 
Eq. (\ref{wave}) into Eq. (\ref{eq:2.1.8}), we obtain 
\begin{eqnarray}
R(r)=A_1J_{\nu}(kr)+A_2N_{\nu}(kr), \label{wave_R}
\end{eqnarray}
where $A_1$ and $A_2$ are constants, $\nu=|m-h|$ and $J_{\nu}$ and $N_{\nu}$ are the Bessel functions of the first kind and the second kind, respectively. The eigen value is given by 
\begin{eqnarray}
\varepsilon_{i}(h)=\frac{\hbar^2 k^2_n(\nu)}{2m_e}, 
\label{eq:4}
\end{eqnarray}
where the index $i$ is for the number of electrons. 
By applying the boundary condition to Eq. (\ref{wave_R}),  
we get a transcendental equation,
\begin{eqnarray}
J_{\nu}(kr_{a})N_{\nu}(kr_{b})-J_{\nu}(kr_{b})N_{\nu}(kr_{a})=0,
\label{eq:2.1.29}
\end{eqnarray}
where 
\begin{eqnarray}
k^2=\frac{2m_e\varepsilon(h) }{\hbar^2}.
\label{eq:3}
\end{eqnarray}
As it is difficult to find solutions analytically, we solve Eq. (\ref{eq:2.1.29}) numerically and can obtain solutions [$k_n(\nu)$, $n=0,1, 2, \cdots$] for $k$, where $n$ are the quantum numbers for $r$.

Under the condition in which $N$ is fixed, we can calculate 
the total energy per sites $E_N(h)$ at $T=0$ 
by 
\begin{eqnarray}
E_N(h)=\frac{1}{N}\sum_{i=1}^{N}\varepsilon_{i}(h). 
\end{eqnarray}
The magnetization at $T=0$ is obtained numerically by 
\begin{eqnarray}
M_{\mu}(h)=-\frac{\partial E_{N}(h)}{\partial h}.
\end{eqnarray}

\section{Results and Discussion}


We show the eigen values [$\varepsilon_{i}(h)$] for $r_b/r_a=10$ in Fig. \ref{fig2}, where we set $\hbar^2/(2m_er_a^2)=1$. 
The eigen values have the period of $\phi_0$, which has been already known\cite{Byers}. 
From a lower figure in Fig. \ref{fig2}, we can see that the eigen values with $1\leq i\leq 7$ have the quantum number of $n=0$. 
The eigen values for $8\leq i\leq 19$ have the quantum number 
of $n=0$ or $n=1$ upon changing $h$, which can be seen in the upper and lower figures in Fig. \ref{fig2}. 
When $r_b/r_a$ is small, the system is close to the one-dimension, and the values of the energy levels (green lines) having $n=1$ 
are larger. Finally, when $r_b/r_a$ is equal to 1, whose case corresponds to the one-dimensional rings, only the energy levels (red lines) belonging to $n=0$ exist. From the upper figure 
in Fig. \ref{fig2}, it is found that $\varepsilon_i(h)$ for 
$20 \leq i\leq 27$ is made by the energy levels 
having to $n=0$, 1 and 2.

The total energy $E_N(h)$ changing $N$ from $1$ to $27$ for $r_b/r_a=10$ are shown in Figs. \ref{fig3} and \ref{fig4}. At $1\leq N\leq 7$, 
$E_N(h)$ oscillates with the period of $\phi_0$ and the phase difference 
between even $N$ and odd $N$ is $\pi$. At $8\leq N\leq 19$, $E_N(h)$ has the period of $\phi_0$ and $\phi_0/2$ for even $N$ and odd $N$, respectively. 
At $N=20, 21, 22, 24, 26$ and 27 
the periods of the oscillations of $E_N(h)$ are $\phi_0$. 
At $N=25$, the period of $E_N(h)$ is $\phi_0/2$. 
Moreover, from an close-up figure for $N=23$ shown 
in Fig. \ref{fig5},the period the oscillations of $E_N(h)$ is almost $\phi_0/3$. We consider that the periodicity of $M_N(h)$ for $20\leq N\leq 27$ is due to the mixing of the levels belonging to $n=0$, 1 and 2 near the Fermi energy.

We show $M_N(h)$ at $1\leq N\leq 27$ for $r_b/r_a=10$ in Figs. \ref{fig6} and \ref{fig7}. 
At $1\leq N\leq 7$, the wave form of $M_N(h)$ is saw-tooth and the phase difference between odd $N$ and even $N$ is $\pi$. We call these a normal saw-tooth pattern (a) and 
a $\pi$-shift saw-tooth pattern (b) for odd $N$ and even $N$, respectively.  
At $8\leq N\leq 19$, the wave form of $M_N(h)$ is saw-tooth and its period is half-period ($\phi_0/2$) for odd $N$. We call it a half-period saw-tooth pattern (c). The wave form of $M_N(h)$ with the period of $\phi_0$ is complicated for even $N$, which is named as a complicated saw-tooth pattern (d), although the wave form of $M_N(h)$ at $N=8$ is not complicated but normal. Since 
at $N=20, 21, 22, 24, 26$ and 27 the wave form of $M_N(h)$ with the period of $\phi_0$ is complicated, those are pattern (d). 
Since $M_N(h)$ at $N=23$ has saw-tooth shape and its period is almost $\phi_0/3$, we call it a almost 1/3-period saw-tooth pattern (e). 
At $N=25$ $M_N(h)$ is classified into pattern (c).

%


\section{Conclusion 
}

In the spinless two-dimensional corbino disk system, we have shown that at $1\leq N\leq 27$ the phase, the period of and saw-tooth wave form of the AB oscillation of magnetization depend on $N$ and can be assorted into the five patterns [normal saw-tooth (a), $\pi$-shift saw-tooth (b), half-period saw-tooth (c), complicated saw-tooth (d) and almost 1/3-period saw-tooth (e)]. In the ideal one-dimensional rings with spin, patterns (d) and (e) have never been shown.

When $N$ is small $(N\leq 7$ for $r_b/r_a=10$), the Fermi energy is always located at the energy levels belonging to $n=0$ even if $h$ is changed. In this case, (i) 
the oscillations become pattern (a) and pattern (b) for odd $N$ and even $N$, respectively. 

When $8\leq N\leq 19$, the Fermi energy is located on the energy levels belonging to either $n=0$ or $n=1$ upon changing $h$. Then, (ii) the oscillations become pattern (c) for odd $N$, which is the same as the one-dimensional rings with spin\cite{Loss} and for even $N$ those become pattern (d), which is not seen in the one-dimensional rings with spin and is different from the complicated saw-tooth shape obtained in hexagonal bilayer rings with spin\cite{Igor}. 

At $20\leq N\leq 27$, the Fermi energy is located on the energy levels belonging to $n=0$ or $n=1$ or $n=2$ when $h$ is changed. Here, 
(iii) there appear pattern (e) in addition to pattern (c) and pattern (d). 

In the Zeeman 
term, $g\mu_BB/2
$, by taking $g=2$ and $\hbar^2/(2m_er_a^2)=1$, we get 
$g\mu_BB/2\simeq 2$ when $h\simeq 1$. 
If $h\gtrsim 1$, the magnitude of the energy by the Zeeman term is very large compared to $\varepsilon_i(h)$ in Fig. \ref{fig2}. 
Namely, if the semiconductor corbino disk with the inner radius of 10 nm is used and the electron number is small ($N\lesssim 30$), the effect of Zeeman splitting due to spin in the disk can be neglected at the high magnetic field ($B\gtrsim 13$ T). 
Therefore, in these conditions, it is expected to confirm (i), (ii) and (iii) experimentally from the magnetization due to the persistent current in semiconductor corbino disk.







\section*{Acknowledgment}
%
We would like to thank Masahiro Hara for useful discussions. 




\end{document}